\documentclass[prc,twocolumn,superscriptaddress,superscriptfootnote]{revtex4}
\usepackage{amssymb}

\usepackage{amsmath,amsfonts,amssymb,epsfig,color}

\newcommand{\SU}[2]{\ensuremath{\mathrm{SU}^{ #1 }( #2 )}}

\newcommand{\SO}[1]{\ensuremath{\mathrm{SO}( #1 )}}

\newcommand{\Spn}[1]{\ensuremath{\mathrm{Sp}( #1 )}}

\newcommand{\su}[2]{\ensuremath{\mathfrak{su}^{ #1 }( #2 )}}

\newcommand{\spn}[1]{\ensuremath{\mathfrak{sp}( #1 )}}

\newcommand{\IAS}{isobaric analog $0^+$ state}
\newcommand{\IASs}{isobaric analog $0^+$ states}
\newcommand{\fpg}{\ensuremath{1f_{5/2}2p_{1/2}
2p_{3/2}1g_{9/2}} }

\newcommand{\flevel}{\ensuremath{1f_{7/2}} }
\newcommand{\dlevel}{\ensuremath{1d_{3/2}} }

\begin{document}

\title{Dynamical symmetry of isobaric analog $0^+$ states in
medium mass nuclei}
\author{K. D. Sviratcheva}
\affiliation{Department of Physics and Astronomy, Louisiana State University,
Baton Rouge, Louisiana 70803, USA}
\author{A. I. Georgieva}
\affiliation{Department of Physics and Astronomy, Louisiana State University,
Baton Rouge, Louisiana 70803, USA}
\affiliation{Institute of Nuclear Research and Nuclear Energy,
Bulgarian Academy
of Sciences, Sofia 1784, Bulgaria}
\author{J. P. Draayer}
\affiliation{Department of Physics and Astronomy, Louisiana State University,
Baton Rouge, Louisiana 70803, USA}
\date{\today}

\begin{abstract}
An algebraic \spn{4} shell model is introduced to achieve a deeper
understanding and interpretation
of the properties of pairing-governed $0^+$ states in medium mass
atomic nuclei. The theory, which
embodies the simplicity of a dynamical symmetry approach to nuclear
structure, is shown to reproduce
the excitation spectra and fine structure effects driven by
proton-neutron interactions and isovector
pairing correlations across a broad range of nuclei.
\end{abstract}


\maketitle

\section{Introduction}
A recent renaissance of interest in pairing correlations in atomic
nuclei is linked to the search for
a reliable microscopic theory for describing the structure of medium
mass nuclei around the $N=Z$ line
where protons and neutrons occupy the same major shell and their
mutual interactions are expected to
strongly influence the structure and decay modes of such nuclei. The revival
of interest in pairing
correlations is also prompted by radioactive beam experiments, which 
are advancing
the exploration of `exotic' nuclei, such as neutron-deficient or
$N\approx Z$ nuclei far off the valley of
stability. Likewise, such a microscopic framework is important for
astrophysical applications, for
example a description of the $rp$-process in nucleosynthesis, which
runs close to the proton-rich side
of the valley of stability through reaction sequences of proton
captures and competing $\beta $ decays
\cite{Langanke98,Schatz98}.

In this paper we show that a simple but powerful group theoretical
approach, with \Spn{4} the
underpinning symmetry, can provide a microscopic description and
interpretation of the properties of
pairing-governed $0^+$ states in the energy spectra of the even-$A$
nuclei with mass numbers $32\le
A\le 100$ where protons and neutrons are filling the same major
shell. In this regard, it is important
to recall that \SO{5} \cite{Hecht,Ginocchio} (with a Lie algebra that
is isomorphic to \spn{4}) has been shown to play a significant role in
the  structure of $fp$-shell $N=Z$ nuclei \cite{EngelLV96}. Indeed, a
model based on this symmetry group can be used to track the
results of an isospin-invariant pairing plus quadrupole shell-model theory
\cite{KanekoHZPRC59}.

A theory that invokes group symmetries is driven by an expectation
that the wave functions of the
quantum mechanical system under consideration can be characterized by
their invariance properties
under the corresponding symmetry transformations. But even if the
symmetries are not exact, if
one can find near invariant operators, the associated symmetries can
be used to help reduce the
dimensionality of a model space to a tractable size. Within the
framework of the \Spn{4} symplectic
group, an approximate symmetry drastically reduces the model space,
which allows the model to be
applied in a broad region of the chart of the nuclides \cite{SGD03}.
This symmetry is adequate only
for a certain class of phenomena, which in our investigation is
related to significant isovector
(isospin $T=1$) pairing correlations in even-$A$ nuclei. While the
model is not valid for all the
states in the energy spectra of the $319$ nuclei we consider (nor can
any model achieve this), it
does yield a realistic reproduction (with only 6 parameters) of the
pairing-governed \IAS ~spectra in medium mass
nuclei where the valence protons and neutrons occupy the same major
shell. The validity and
reliability of the model with respect to the interactions it includes
are confirmed additionally via
a finite energy difference method that we employed to reproduce
detailed nuclear structure, including
$N=Z$ anomalies, isovector pairing gaps and staggering effects \cite{SGD03stg}.

The present investigation shows the advantage of the algebraic
\spn{4} approach over other theoretical
studies.  Namely, the \Spn{4} model, which is based on Helmer's
quasi-spin scheme \cite{Helmers}, is
both simpler and provides a better understanding of the fundamental
nature of the nuclear interaction
compared, for example, to the more general but also more elaborate
$U(4\Omega)$ model
\cite{TalmiSimpleModels} based on the conventional seniority scheme
of Racah and Flowers
\cite{Racah,Flowers}. It also yields a better and more detailed
microscopic description of isovector
pairing correlations and proton-neutron interactions than mean-field
theories and results extracted
from semi-empirical mass formulae. In addition, it can be used in
higher-lying shells where other
approaches cannot be applied.

\section{Pairing models: from \SU{}{2} to \Spn{4}}

It has been recognized for a long time that ground states of
even-even nuclei reflect strongly on
the nature of the nuclear interaction, especially its propensity to
form correlated, angular momentum
$J=0$ pairs \cite{BohrMottelsonPines, BohrMottelson}. It is also well
known that the low-lying energy
spectrum of isotopes of a doubly magic core (such as $^{40}$Ca or
$^{56}$Ni) can be well reproduced in
terms of neutron pair [$nn$] addition to the spectrum of states of
the core \footnote{Without
compromising the theory, one can consider closed shells as part of an inert
core that is spherical and does not
affect directly the single-particle motion of the valence nucleons in
the last unfilled shell.}.  In
complete analogy with this, if one adds to the ground state of the
core $J=0,\ T=1$ pairs of nucleons
(two protons $pp$, two neutrons $nn$, or a proton and a neutron
$pn$), one would construct fully paired
states that in general reflect the close interplay of like-particle
and $pn$ pairs. Provided that the
isospin is (almost) a good quantum number, which is typically the
case for low-lying states in light
and medium mass even-$A$ nuclei with valence protons and neutrons
simultaneously filling the same major
shell, fully paired $0^+$ states built in this way describe
isobaric analog $0^+$ states ({\it
IAS}) in nuclei across the entire shell.
In the mass range $32 < A < 100$
where the influence of shape deformation on these $0^+$ {\it IAS}
states is relatively weak, this notion
of fully paired (seniority zero) states is a valid, albeit approximate
picture. While in even-even
nuclei within this region the ground states are such fully paired
$0^+$ states, this it is not always
the case for odd-odd nuclei.  The strong proton-neutron interaction
usually drives the state with the
least  symmetry energy ($\sim T^2$, where $T$ is nuclear isospin)
lowest. However, it is not the
purpose of this article to investigate such states. Rather, we
consider the isobaric analog $0^+$
state, which is typically higher in energy and is strongly influenced
by isovector pairing
correlations. Even though the states are represented as $J=0$ pairs,
the interaction that governs them
is not exclusively pairing in the $J=0$ channel, but must also
include the important $pn$
$J_{\text{odd}}\geq 1$ isoscalar ($T=0$) interaction.  The
significant interplay between these
isovector and isoscalar interactions is evident in the low-lying
structure of $N=Z$ odd-odd nuclei and
has been the focus of a large number of experimental
\cite{ZeldesLirin76,Rudolph,Vincent98,GNarro} and
theoretical studies \cite{CivitareseReboiroVogel, LangankeDKR,
Dean97, SatulaDGMN, PovesMartinezPinedo,
MartinezPinedoLV, KanekoHZPRC59, SatulaWyss, MacchiavelliPLB,
Macchiavelli00, Vogel00}.

The zero seniority $0^+$ states can be constructed as ($T=1$)-paired fermions
\begin{equation}
\left|n_{+1},n_{0},n_{-1}\right) =\left( \hat{A}^{\dagger
}_{+1}\right)  ^{n_{+1}}\left(
\hat{A}^{\dagger }_{0}\right) ^{n_{0}}\left( \hat{A}^{\dagger
}_{-1}\right) ^{n_{-1}}\left|
0\right\rangle ,
\label{GencsF}
\end{equation}
where $n_{+1,0,-1}$ are the numbers of $J=0$ pairs of each kind, $pp$,
$pn$, $nn$, respectively, and $\left| 0\right\rangle$ denotes the vacuum state.
The transition operator, which changes the number of particles in a
pairwise fashion,
$\hat{A}^{\dagger }_{0,+1,-1}$, creates a proton-neutron ($pn$) pair,
a proton-proton
($pp$) pair or a neutron-neutron $(nn)$ pair of total angular momentum
$J^{\pi}=0^+$ and isospin $T=1$. Each operator $\hat{A}^{\dagger
}_\mu ,\ \mu =0,+1,-1$, together
with its conjugate pair-annihilation operator, $\hat{A}_\mu $, and a
pair number operator generate an \SU{\mu }{2} subgroup of \Spn{4}, which
in the case of $\mu =\pm
$ is the standard like-particle pairing Kerman's \SU{}{2} group
\cite{Kerman}.

 From a microscopic perspective, the pair-creation (pair-annihilation)
operators,
$\hat{A}^{(\dagger )}$, are realized in terms of creation $c
_{jm\sigma }^\dagger$ and
annihilation $c _{jm\sigma }$ single-fermion operators with the standard
anticommutation relations
$\{c _{jm\sigma },c _{j^{\prime  }m^{\prime }\sigma ^{\prime }}^{\dagger
}\}=\delta _{j,j^{\prime }}\delta _{m,m^{\prime }}\delta _{\sigma ,\sigma
^{\prime }},$ where these operators create (annihilate) a particle of type
$\sigma =\pm 1/2$ (proton/neutron) in a state of total angular momentum $j$
(half integer) with projection
$m$ in a finite space $2\Omega =\Sigma _j (2j+1)$. There are ten independent
scalar products (zero total angular momentum) of the fermion operators:
\begin{eqnarray}
       \hat{A}^{\dagger }_{\mu =\sigma+\sigma^{\prime}}&=&
\frac{1}{\sqrt{2\Omega (1+\delta_{\sigma\sigma ^{\prime}})}}
\sum_{jm} (-1)^{j-m} c_{jm\sigma}^\dagger c_{j,-m,\sigma
^{\prime}}^\dagger,\nonumber \\
       \hat{A}_\mu &=& (\hat{A}^{\dagger }_\mu)^\dagger,\nonumber \\
\hat{T}_\pm &=& \frac{1}{\sqrt{2\Omega}} \sum_{jm} c^\dagger_{jm,\pm
1/2}  c_{jm,\mp 1/2}, \nonumber \\
\hat{N}_{2\sigma }&=& \sum_{jm} c^\dagger_{jm\sigma }  c_{jm\sigma },
\label{gen}
\end{eqnarray}
which form a fermion realization of the symplectic \spn{4} Lie algebra. Such an
algebraic structure is exactly the one needed to describe isovector
(like-particle plus $pn$) pairing correlations and isospin symmetry
in nuclear \IASs. In
(\ref{gen}),
$\hat{N}_{\pm 1}$ are the valence proton (neutron) number operators. The
generators $\hat{T}_{0}$ and $\hat{T}_{\pm}$ are associated with the
components of the isospin
of the valence particles and close on an \su{T}{2} subalgebra of \spn{4}. In
terms of the generators of the \Spn{4} group (\ref{gen}), the
operator that counts the total number of
valence particle $n$ is expressed as
$\hat{N}=\hat{N}_{+1}+\hat{N}_{-1}$ and the third isospin
projection operator is
$\hat{T}_{0}=(\hat{N}_{+1}-\hat{N}_{-1})/2$.

While the \Spn{4} symplectic group embeds in itself
the well-known symmetry of like-particle pairing,  $\Spn{4} \supset
SU^{+}(2) \otimes SU^{-}(2)$, it brings into the theory the significant
interaction between protons and neutrons through the reduction chains
$Sp(4)\supset U^{\mu }(2)\supset U^{\mu }(1)\otimes SU^{\mu }(2)$ with
$\mu =0$ (proton-neutron pairing symmetry) and $\mu =T$ (isospin
symmetry). These group reductions allow the \Spn{4}-invariant
degenerate energy states to
split, which is the case of physical interest. Such a dynamical
symmetry that our model
possesses provides for a natural classification scheme of  nuclei as
belonging to a
single-$j$ level or a major shell (multi-$j$), which are mapped to
the algebraic multiplets. This classification also extends to the
corresponding ground and  excited states of the nuclei.

The general model Hamiltonian with \Spn{4} dynamical symmetry consists of
one- and two-body  terms and can be expressed through the
\Spn{4} group generators,
\begin{eqnarray}
H =&-G\sum _{i=-1}^{1}\hat{A}^{\dagger }_{i}
\hat{A}_{i}-F \hat{A}^{\dagger }_{0}\hat{A}_{0}-\frac{E}{2\Omega} (\hat{T}
^2-\frac{3\hat{N}}{4 })
\nonumber \\
&-D(\hat{T}
_{0}^2-\frac{\hat{N}}{4})-C\frac{\hat{N}(\hat{N}-1)}{2}-\epsilon
\hat{N},
\label{clH}
\end{eqnarray}
where $\hat{T}^2=\Omega \{ \hat{T}_+,\hat{T}_-\}+\hat{T}_0^2$ is the
isospin operator, $G,F,E,D$ and
$C$ are interaction strength parameters  and
$\epsilon >0$ is the Fermi level energy.
This Hamiltonian conserves the number of
particles ($n$) and the third projection ($T_0$) of the isospin, 
while it includes
scattering of a $pp$ pair and  a $nn$ pair into two $pn$ pairs and 
vice versa. 

As we have shown in \cite{SGD03}, the algebraic model 
Hamiltonian (\ref{clH}) arises naturally within a microscopic picture.
Using relations (2), it can be rewritten in standard second quantized
form, which in turn defines the physical nature of the interaction
and its strength. For example, in this way, one can identify the
parameters $G/\Omega$ and $(G+F)/\Omega$ in (\ref{clH}) with the strength
of the $J=0$ $T=1$ pairing interaction between two protons (neutrons)
and  a proton and a neutron, respectively. The $C$, $D$, and
$E$ parameters are related to the expectation value of an average
$J$-independent interaction, which includes for example high-$J$
like-particle interaction with a strength specified by the parameters
$C+\frac{D}{2}+\frac{E}{4\Omega }$\cite{SGD03}.

Furthermore, the $E$ term in (\ref{clH}), together with the $C$
term,  is related to the
microscopic nature of the odd $J$ isoscalar ($T =0$) interaction 
between a proton and a
neutron, $-\frac{E}{2\Omega} (\hat{T}
^2-\frac{3\hat{N}}{4 })-C\frac{\hat{N}(\hat{N}-1)}{2}=-\frac{E}{2\Omega}(
\hat{T}^2-\frac{\hat{N}}{2}-
\frac{\hat{N}^2}{4})-(C+\frac{E}{4\Omega})\frac{\hat{N}(\hat{N}-1)}{2}$, 
where the first part
comprises the $J$-independent $pn$ isoscalar force. It is diagonal in 
the isospin basis and
can be compared to \cite{HasegawaKanekoPRC59,KanekoH99}. In addition, the
quadratic in $\hat{N}$ term can be understood as an
average two-body  interaction between the
valence particles (note that for $n$ equivalent particles there are
$\binom{n} {2}=\frac{n(n-1)}{2}$ particle couplings).

 From another perspective, the $E$-term can be related to the symmetry 
energy \cite
{Hecht,TalmiSimpleModels} as its expectation value in states with 
definite isospin is of the
form $T(T+1)$, which enters as a symmetry term in many nuclear mass 
relationships
\cite{JaneckeB74,DufloZuker}.
We refer to the $E$-term as {\it a symmetry term}, although it is 
common to address the
symmetry energy in a slightly different way: the
$T (T +1)$ term together with the isospin dependence of the isovector 
pairing term
yield both symmetry  ($\sim T ^2\sim (Z-N)^2$) and
Wigner ($\sim T $) energies \cite{Wigner37}. The first one was 
originally included in the
Bethe-Weizs\"acker semi-empirical mass formula 
\cite{Weizsacker35,Bethe36} and implies that
the nuclear symmetry energy has the tendency toward stability for 
$N=Z$. The Wigner energy is
associated with proton-neutron exchange interactions and is 
responsible for a sharp energy
cusp at $N=Z$ leading to an additional binding of self-conjugate 
nuclei \cite{BohrMottelson}.
In short, the symmetry energy together with the terms that are linear 
in $\hat{N} $ (\ref{clH}) can
be directly related to a typical mass formula 
\cite{Weizsacker35,Bethe36}, while in addition our model
improves the description of isovector pairing correlations and high 
$J$ identical-particle and $pn$
interactions and uses an advanced Coulomb repulsion correction
\cite{RetamosaCaurier}.

In this way, Hamiltonian (\ref{clH}) includes an isovector ($T=1$)
$nn$, $pn$, and  $pp$ pairing interaction ($G\geq 0$ for attraction) 
and a diagonal
(in an isospin basis) isoscalar ($T=0$) proton-neutron force, which 
is related to
the so-called symmetry term ($E$). Hence, the model Hamiltonian 
(\ref{clH}) includes the
dominant interactions that govern the $0^+$ states under 
consideration and provides for an
exact solution of the present problem.

In addition, the  $D$-term in (\ref{clH}) introduces isospin symmetry 
breaking and the $F$-term accounts for a plausible, but weak,
isospin mixing. While both terms
are significant in the investigation of certain types of phenomena 
\cite{SGD04IM},
the study of their role is
outside the scope of this paper. These parameters yield quantitative 
results that are
better than the ones with $F=0$ and $D=0$: for example, in the case of the
\flevel level the variance between the model and experimental 
energies of the lowest
\IASs~ increases by $85\%$ when the $D$ and $F$ interactions are 
turned off. At the same
time, the latter provide only fine adjustments compared to the main 
driving forces
incorporated in (\ref{clH}). In this sense, a simpler isospin
invariant $SO(5)$ model is suitable for a
qualitative description of the \IAS~ energy spectra of nuclei.

\section{Description of \IASs}

\subsection{Interaction strength parameters}

The interaction strength parameters are estimated in a fit of the minimum
eigenvalue $(-E_0)$ of the $H$ energy operator (\ref{clH}) to the 
Coulomb corrected
\cite{RetamosaCaurier} experimental energies 
\cite{AudiWapstra,Firestone} of the lowest
isobaric analog $0^+$ states of even-$A$ nuclei (ground states for 
even-even  nuclei and some
[$N\approx Z$] odd-odd nuclei).
We use the Coulomb correction $V_{Coul}(A,Z)$ derived in
\cite{RetamosaCaurier} so that the Coulomb corrected energies are 
adjusted to be
$E_{0,exp}(A,Z)=E^C_{0,exp}(A,Z)+V_{Coul}(A,Z)-E_{0,exp}(A_{c},Z_{c})$, where
$E^C_{0,exp}(A,Z)$ is the total measured (positive) energy including 
the Coulomb energy and
$E_{0,exp}(A_{c},Z_{c})=E^C_{0,exp}(A_{c},Z_{c})+V_{Coul}(A_{c},Z_{c})$
is the corrected energy of a nuclear core.
Analogously, the theoretically predicted energies that include the Coulomb
repulsion can be obtained as 
$E_{0}^C(A,Z)=E_{0}(A,Z)-V_{Coul}(A,Z)+E_{0,exp}(A_{c},Z_{c})$. 

\begin{table}[th]
\caption{Parameters and statistics for three regions ({\bf I},
{\bf II}, and {\bf III})
specified by the valence model space. $G$, $F$, $C$, $D$, $E$, $\epsilon $,
and $\chi $ are in MeV, $SOS$ is in MeV$^{2}$.}
\center{
\begin{tabular}{cccc}
\hline
\hline
\text{Parameters}  & {\bf I} & {\bf II}  & {\bf III} \\
\text{and }        & $(\dlevel )$ & $(\flevel )$  &  $(1f_{5/2}2p_{1/2}$ \\
\text{statistics}  &       &        & $2p_{3/2}1g_{9/2})$ \\
\hline
$G/{\Omega }$      & 0.702 &  0.453 &  0.296  \\
$F/{\Omega }$      & 0.007 &  0.072 &  0.056  \\
$C$                & 0.815 &  0.473 &  0.190  \\
$D$                & 0.127 &  0.149 & -0.307  \\
$E/{(2\Omega )}$   &-1.409 & -1.120 & -0.489  \\
$\epsilon $        & 9.012 &  9.359 &  9.567  \\
\hline
$SOS$                & 1.720 & 16.095 &300.284  \\
$\chi $            & 0.496 & 0.732  &  1.787  \\
\hline \hline
\end{tabular}
}
\label{tab:fitStat}
\end{table}
The fitting procedure was performed separately for  three groups of nuclei
with valence nucleons occupying ({\bf I}) the
$1d_{3/2}$ level with a
$^{32}S$ core, ({\bf II}) the
$1f_{7/2}$ level with a $^{40}Ca$ core, and ({\bf III}) the \fpg
shell with a $^{56}Ni$
core \cite{SGD03}. The results reveal that the model interaction
accounts quite well for the
available experimental energies for a total of $149$ nuclei (refer to
the small value of the
$\chi $-statistics in Table \ref{tab:fitStat}, where $\chi ^2$ is the
sum of squares, $SOS$,
divided by the difference between the number of data cases and the
number of fit parameters)
\cite{SGD03}. The values of the parameters
in $H$ (\ref{clH}) are  kept fixed hereafter (Table \ref{tab:fitStat}).

In the case of the \fpg  shell ({\bf III}), the parameters
of the effective interaction in the \Spn{4} model with degenerate
multi-$j$ levels are likely
to be influenced by the non-degeneracy of the single-particle orbits.
Nevertheless, as the dynamical
symmetry properties of the two-body interaction in nuclei from this
region are not lost, the model
remains a good multi-$j$ approximation \cite{SGD03}, which is
confirmed with the use of
various discrete derivatives of the energy function \cite{SGD03stg}.

\begin{figure}[h]
\centerline{\epsfxsize=3.4in\epsfbox{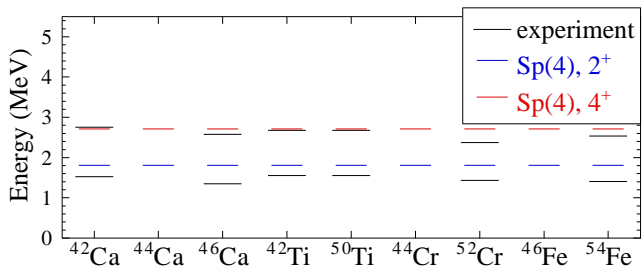}}
\caption{(Color online) Low-lying energy spectra of near closed-shell
nuclei in the \flevel level as
described by the like-particle pairing limit of the \Spn{4} model and
compared to the experiment.}
\label{pairLimitEnSpectra}
\end{figure}
\begin{figure}[h]
\centerline{\epsfxsize=3.4in\epsfbox{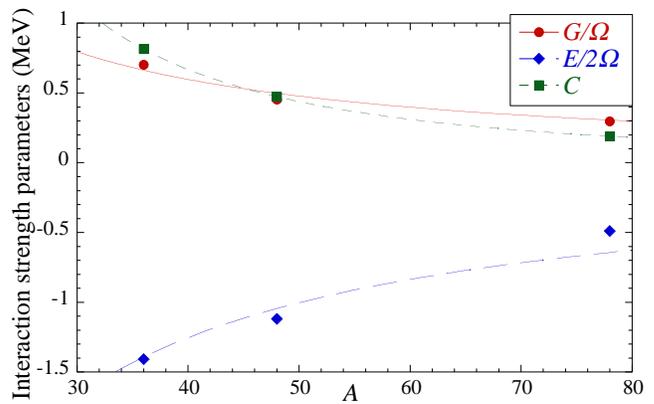}}
\caption{(Color online) Dependence of the interaction strength
parameters $G/\Omega $, $E/2\Omega $,
and $C$ (in MeV) on the mass number
$A$ with values from the three regions, ({\bf I}), ({\bf II}) and
({\bf III}).}
\label{ParamFnA}
\end{figure}

While the optimum fit is statistically determined solely by $\chi $,
the physical aspect of the
nuclear problem requires, in addition, the estimate for the
parameters to be physically valid.
Indeed, the values of the like-particle pairing strength $G$,
obtained by our \Spn{4} model, yield
consistent results \cite{SGD03stg} with the experimental pairing gaps
derived from the odd-even mass
differences \cite{NilssonP61,RingS80}. In this way, the $G$ values
are expected to
reproduce the low-lying vibrational spectra of near closed-shell
nuclei in the \SU{\pm
}{2} limit of the model (like-particle pairing) (Figure
\ref{pairLimitEnSpectra}). When the results
from the three nuclear regions ({\bf I}, {\bf II}, and {\bf III}) are
considered, the pairing
strength parameter is found to follow the well-known $1/A$ trend \cite
{KermanLawsonMacfarlane,KisslingerSorensen,Lane,BohrMottelson,DudekMS},
\begin{eqnarray}
\frac{G}{\Omega }&=\frac{23.9 \pm 1.1}{A},\quad &R^2=0.96,
\end{eqnarray}
where $R^2$ is a coefficient of correlation and represents the
proportion of variation in the strength parameter accounted for by
the analytical
curve (Figure \ref{ParamFnA}).
Similarly, the values of the other strength parameters lie on a curve
that decreases with
nuclear mass $A$ (Figure \ref{ParamFnA})
\begin{equation}
\begin{array}{cclc}
\frac{E}{2\Omega }&=&\frac{-50.2 \pm 3.3}{A}, &R^2=0.93,\\
C&=&\left( \frac{32.30 \pm 0.02}{A} \right)^{1.887\pm 0.004}, &R^2=0.99.
\end{array}
\end{equation}
As expected for a symmetry energy term, the $1/A$ dependence 
holds for the parameter
$E$. The dependence of $C$ on the mass number $A$ suggests that the 
quadratic correction
$C\frac{n(n-1)}{2}$ to the mean field may change slowly from one
nucleus to another, which is consistent with the saturation of the 
nuclear force. Although the data set used in the fitting procedure was
rather small, the trend toward a smooth
functional dependence of the interaction strength parameters on the 
mass number $A$ reveals their
{\it global } character, namely the interactions in the model 
Hamiltonian (\ref{clH}) are related to
an overall behavior common to all nuclei.

Furthermore, the Wigner energy \cite{Wigner37}, $-W2T $, is
implicitly included in the
\Spn{4} theoretical energy, which in turn makes the estimation of its
strength possible.
The Wigner energy appears as the term that is the linear in $T $ in
the $pn$ isoscalar force
(proportional to the symmetry term) and in the isovector pairing through the
second-order Casimir invariant of \spn{4}. In a good-isospin regime,
the symmetry energy contribution
is $-\frac{E}{2\Omega }T(T+1)$  [due to the $\hat{T}^2$-term in
(\ref{clH})], and the $W$ interaction
strength parameter can be expressed through the model parameters
(Table \ref{tab:fitStat}) as
$W=\frac{E-G}{4\Omega }$. In the framework of the
\Spn{4} model, the estimated values for $W$ from the three regions
({\bf I}, {\bf II} and {\bf
III}) are found to lie on a curve
\begin{equation}
W=\frac{-31 \pm 2}{A},\ R^2=0.96,
\end{equation}
with a very good correlation coefficient $R^2$ and a remarkably close
value to most other estimates:
$W=-30/A $
\cite{MollerNK97} , $W=-37/A $ \cite{DufloZuker}, $W=-37.4/A $
\cite{KanekoH99}, $W=-42/A $
\cite{MyersS96} and $W=-47/A $ \cite{Vogel00}.

In short, the outcome of the optimization procedures shows that the
effective interaction
with \Spn{4} dynamical symmetry provides a reasonable description of
the lowest \IASs, retaining the
physical meaning and validity of its microscopic nature.

\subsection{Energy spectra of the isobaric analog $0^+$ states}

\begin{figure}[t]
\centerline{\epsfxsize=3.5in\epsfbox{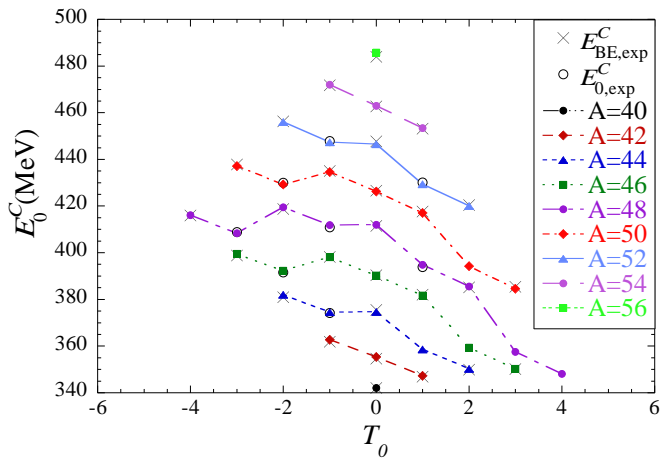}}
\caption{(Color online) Isobaric analog $0^{+}$ state energy,
$E^C_{0}$, in MeV (including the
Coulomb energy) versus the isospin projection $T_0$ for the isobars
with $A=40$ to $A=56$ in the
\flevel level, $\Omega _{\frac{7}{2}}=4$. The experimental binding energies
$E^C_{\text{BE,exp}}$ (symbol ``$\times
$'') are distinguished from the experimental energies of the isobaric analog
$0^{+}$ excited states $E^C_{0,\exp }$ (symbol ``$\circ $''). Each
line connects
theoretically predicted energies of an isobaric sequence. }
\label{CaE0expTh}
\end{figure}

In all three nuclear regions, there is good agreement with
experiment (small $\chi $-statistics), as can be seen
in Figure \ref{CaE0expTh} for the isobars $A=40-56$ in the
\flevel level $({\bf
II})$. The theory predicts the lowest \IAS~energy of nuclei
with a deviation ($\chi /\Delta E_{0,exp}\times 100 [\% ]$) of
$0.7\%$ for $({\bf
I})$ and $0.5\%$ for $({\bf II})$ and $({\bf III})$ in the corresponding energy
range considered, $\Delta E_{0,exp}$.

\begin{widetext}

\begin{figure}[h]
\centerline{\epsfxsize=4.4in\epsfbox{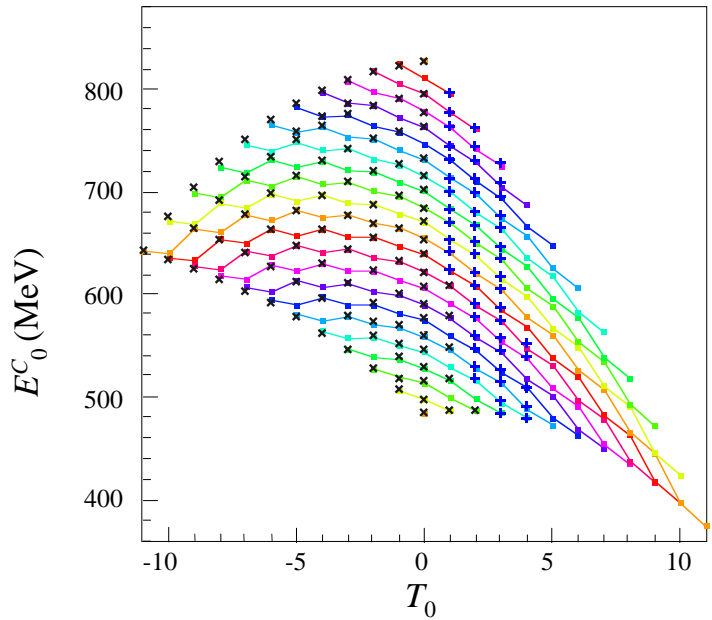}}
\caption{(Color online) Theoretical energies $E^C_0$ (including the
Coulomb energy contribution) of the
lowest \IASs ~ for isobars (connected by lines in different
colors) with mass number
$A=56,58,\dots ,100$  in the \fpg major shell ($^{56}$Ni core), compared to
experimental values (black `$\times $') and semi-empirical estimates in
\cite{MollerNK97} (blue `$+$'). }
\label{NiE0expTh}
\end{figure}

\end{widetext}

The fitting procedure not only estimates the magnitude of the interaction
strength and determines how well the model Hamiltonian ``explains" the
experimental data, it also can be used to predict nuclear energies
that have not
been measured. This includes energies of nuclei with odd number of
protons and neutrons and as
well nuclei away from the valley of stability with $N\approx Z$ or
proton-rich that are of great
interest in modern astrophysical studies. From the fit for the
\flevel case, the binding energy of the proton-rich $^{48}$Ni nucleus
is estimated
to be $348.19$ MeV, which is $0.07\%$ greater than the sophisticated
semi-empirical
estimate of \cite{MollerNK97}. Likewise, for the odd-odd nuclei that
do not have measured
energy spectra the theory  can predict the energy of their lowest
isobaric analog $0^{+}$ state: $358.62$ MeV ($^{44}$V), $359.34$ MeV
($^{46}$Mn),
$357.49$ MeV ($^{48}$Co), $394.20$ MeV ($^{50}$Co) (Figure
\ref{CaE0expTh}). The
\Spn{4} model predicts the relevant $0^+$ state energies for an additional 165
even-$A$ nuclei in the medium mass region ({\bf III}) plotted in Figure
\ref{NiE0expTh}. The binding energies for 25 of them are also calculated in
\cite{MollerNK97}. For these even-even nuclei, we predict binding energies that
on average are $0.05\%$ less than the semi-empirical approximation
\cite{MollerNK97}.

\begin{figure}[th]
\centerline{\epsfxsize=3.0in\epsfbox{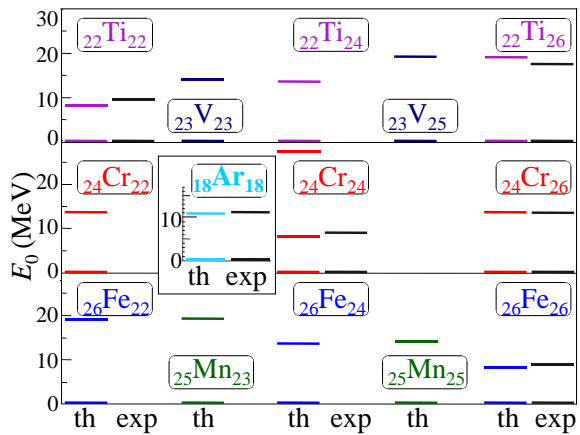}}
\caption{(Color online) Theoretical and experimental (black lines)
energy spectra of the higher-lying pairing-governed \IASs~ for isotopes in the
\flevel shell ($^{40}Ca$ core). Insert: First excited \IAS~
energy in  $^{36}$Ar in the \dlevel shell ($^{32}S$ core) in
comparison to its experimental
value.}
\label{enSpectraCa}
\end{figure}

Without varying the values of the interaction strength parameters (Table
\ref{tab:fitStat}), the energy of the higher-lying pairing-governed \IASs~ in
nuclei  under consideration can be
theoretically calculated. These states are eigenvectors of the model
Hamiltonian (\ref{clH}) and
differ among themselves in their pairing modes due to the close
interplay between like-particle and
$pn$ pairs. The theoretical energy spectra of these \IASs~ agree
remarkably well with the available
experimental values
\footnote{The energy spectra of nuclei in the ({\bf III}) region with
nuclear masses $56<A<100$ is
not yet completely measured, especially the higher-lying $0^+$
states. This makes a
comparison of the theory to the experiment impossible.}
(Figure~\ref{enSpectraCa}). This agreement, which is observed not
only in single cases but throughout
the shells, represents a valuable result. This is because the
higher-lying $0^+$ states under consideration
constitute an experimental set independent of the data that enters
the statistics to determine the model parameters in (\ref{clH}). Such
a result is,
first, an independent test of the physical validity of the strength parameters,
and, second, an indication that the interactions interpreted by the
\Spn{4} model
Hamiltonian are the main driving force that defines the properties of these
states. In this way, the \Spn{4} dynamical symmetry of the zero-seniority
$IAS$ $0^+$ states of  even-$A$ nuclei reveals
a simple and fundamental aspect of the nuclear interaction related to
isovector $J=0$ pairing
correlations and higher-$J$ proton-neutron interactions. Moreover, the simple
\Spn{4} model can be used to provide a reasonable prediction of the
(ground and/or excited) pairing-governed \IASs~
in proton-rich nuclei with energy spectra not yet experimentally
fully explored.

Such a conclusion is based furthermore on our complementary
investigation \cite{SGD03stg} on the fine
structure phenomena among the \IASs. A study of this kind is quite
necessary because it is
well-known that a good reproduction of the experimental nuclear
energies does not guarantee straight
away agreement of the fine structure of nuclei in comparison to the
experiment. We have examined such detailed features by discrete
approximations of derivatives of the
energy function (\ref{clH}) filtering out the strong mean-field influence
\cite{SGD03stg}. In short, this investigation revealed a remarkable
reproduction of the two-proton
and two-neutron separation  energies, the irregularities found around
the $N=Z$ region, the
like-particle and $pn$ isovector pairing gaps, the significant role
of the symmetry energy and
isovector pairing correlations in determining the fine nuclear
properties, and a prominent
staggering behavior observed between groups of even-even and odd-odd nuclides
\cite{SGD03stg}. This
study confirmed additionally the validity and reliability of the
group theoretical \Spn{4} model and
the interactions it includes.

\section{Conclusions}
In this paper we presented a simple \Spn{4} model that achieved a
reasonable prediction of the pairing-governed
\IAS ~energy spectra of a total of 319 even-even and odd-odd nuclei 
with only six
parameters. The model
Hamiltonian is a two-body effective interaction, including
proton-neutron and like-particle pairing
plus symmetry terms (the latter is related to a proton-neutron
isoscalar force). We compared
the theoretical results with experimental values and examined in detail their
outcome. While the model describes only the pairing-governed \IASs 
~of even-even
medium mass nuclei with protons and
neutrons occupying the same shell, it reveals a fundamental feature
of the nuclear interaction, which
governs these states. Namely, the latter possess clearly a simple
\Spn{4} dynamical symmetry.
Such a symplectic \Spn{4} scheme allows also for an extensive
systematic study of various
experimental patterns of the even-$A$ nuclei.

\section*{Acknowledgments}

\vskip .5cm This work was supported by the US National Science
Foundation, Grant
Number 0140300. The authors thank Dr. Vesselin G. Gueorguiev for his
computational {\small MATHEMATICA}\ programs for non-commutative algebras.

\end{document}